\documentclass[12pt,english,reqno]{smfart}
\usepackage{mathptm}
\theoremstyle{plain}
\newtheorem{thm}{Theorem}[section]
\newtheorem{lem}[thm]{Lemma}

\theoremstyle{definition}
\newtheorem*{rem}{Remark}
\renewcommand{\theequation}{\arabic{section}.\arabic{equation}}
\begin{document}
\title[3D hyperbolic sigma model]
{Spontaneous symmetry breaking of a hyperbolic sigma model in
three dimensions}
\author{T.~Spencer}
\address{School of Mathematics, Institute for Advanced Study,
Princeton N.J., USA}
\author{M.R.~Zirnbauer}
\address{Institut f\"ur Theoretische Physik, Universit\"at zu
K\"oln, Germany}
\date{September 21, 2004}
\email{spencer@math.ias.edu, zirn@thp.uni-koeln.de}
\begin{abstract}
Non-linear sigma models that arise from the supersymmetric approach to
disordered electron systems contain a non-compact bosonic sector.  We
study the model with target space $\mathrm{H}^2$, the two-hyperboloid
with isometry group $\mathrm{SU}(1,1)$, and prove that in three
dimensions moments of the fields are finite in the thermodynamic
limit. Thus the non-compact symmetry $\mathrm{SU}(1,1)$ is
spontaneously broken.  The bound on moments is compatible with the
presence of extended states. \\

\end{abstract}
\maketitle

\section{Introduction}

Random-matrix ensembles such as the Gaussian Unitary Ensemble
(GUE) and its cousins have attracted much attention in both the
physics and mathematics community because of its many connections
to statistical many-body theory, integrable systems, number theory
and probability. This article is motivated by the study of
Gaussian matrix ensembles which incorporate spatial structure and
thus are no longer mean field in character.  These ensembles are
sometimes called band GUE models. They have the advantage of being
mathematically more tractable than say random Schr\"odinger
operators and yet they are expected to share the same qualitative
features.

About twenty-five years ago, Wegner \cite{wegner79,sw} introduced
hyperbolic non-linear sigma models to study band GUE models and
disordered electron systems with $N$ orbitals per site. In the
simplest case of these sigma models the hyperbolic `spins' indexed
by lattice sites of $\mathbb{Z}^d$ take values in the hyperbolic
plane $\mathrm{H}^2$ equipped with its $\mathrm{SU}
(1,1)$-invariant geometry.  Soon thereafter, Efetov \cite{efetov}
extended Wegner's work and introduced a class of supersymmetric
non-linear sigma models.

The supersymmetric formalism has the advantage of making it
possible to perform the disorder average and rigorously convert
random matrices to a problem in statistical mechanics.  In
particular, averages of products of Green's functions become
statistical mechanical correlation functions.  The resulting
problem is how to analyse such statistical mechanics systems. One
of the main difficulties in this analysis is the non-compact
hyperbolic symmetry identified by Wegner which naturally arises
when studying spectral and transport properties of disordered
systems. This paper is devoted to showing that a certain class of
$\mathrm{SU}(1,1)$ sigma models can be effectively analysed in
three dimensions by using horospherical coordinates and
Brascamp-Lieb inequalities.

Let $\Lambda$ be a periodic box in $\mathbb{Z}^d$ (centered at 0)
with volume $|\Lambda|$ and define $R(i,j)$ with $i, j \in
\Lambda$ to be the elements of a Hermitian matrix drawn from the
GUE.  Thus the probability density is taken to be proportional to
$\exp (- \mathrm {Tr} \, R^2) \prod dR(i,j)$.  Now let $J$ be a
symmetric matrix with positive entries $J(i,j)$ which are small
when the distance $|i-j|$ is large. Then define a band matrix $H$
with matrix elements
\begin{equation}
  H(i,j) = \sqrt{J(i,j)} \, R(i,j) \;.
\end{equation}
If we set
\begin{equation}\label{1.2}
  J(i,j) = \big( - W^2 \Delta + 1 \big)^{-1}(i,j)
\end{equation}
(with $\Delta$ the Laplacian of the lattice $\Lambda$), then this
corresponds to a band of width $W$. Note that $J(i,j)$ has
exponential decay $\mathrm{e}^{- |i-j| / W}$. Another convenient
choice of $J$ is given as follows. Suppose that $\Lambda$ is tiled
by identical cubes of width $W$. Then define
\begin{equation}\label{1.3}
  J(i,j) = \left\{
  \begin{array}{ll}
    J_0 &\text{if $i$ and $j$ belong to the same cube,} \\
    J_1 &\text{if $i$ and $j$ belong to adjacent cubes,} \\
    0 &\text{otherwise}.
  \end{array} \right.
\end{equation}
In both cases the matrix elements far from the diagonal are
suppressed or vanish, and now the dimension of the lattice plays
an important role. One expects that $H$ has qualitatively the
features of a random Schr\"odinger operator as we let the box
$\Lambda$ approach $\mathbb {Z}^d$. In fact, for the
infinite-volume limit and fixed $W$ we know that $H$ has pure
point spectrum (and thus localization) at all energies in one
dimension, and in any dimension there is localization for energies
in the Lifshitz tails.  Large $W$ is expected to be roughly
inversely proportional to the strength $\lambda$ of the random
potential. For example in one dimension the localization length is
proportional to $\lambda^{-2}$ for the random Schr\"odinger
operator, and proportional to $W^2$ for a band random matrix
\cite{FyoMir}.

For dimension $d = 3$, with $J$ given by (\ref{1.2}) and large
$W$, the average local density of states
\begin{equation}\label{1.4}
  \rho(E) = \pi^{-1} \mathfrak{Im} \left\langle \big( H - E -
  \mathrm{i} \varepsilon \big)^{-1}(x,x) \right\rangle
\end{equation}
was studied using the supersymmetric formalism in the limit when
$\varepsilon$ goes to zero.  $\rho(E)$ was shown to be smooth for
$E$ in an interval around zero and field correlations were proven
to decay exponentially fast \cite{DPS}.

To get information about time evolution or the behavior of the
eigenstates, expectations like (\ref{1.4}) are not sufficient.
Instead one must consider
\begin{equation}\label{1.5}
  \left\langle \big| ( H - E - \mathrm{i} \varepsilon )^{-1}(x,y)
  \big|^2 \right\rangle
\end{equation}
and study its behavior as $\varepsilon$ goes to zero. If
(\ref{1.5}) remains bounded for $x = y$ and $\varepsilon |\Lambda|
\rho(E) = 1$, then the eigenstates near $E$ are extended in the
sense that the $L^4$ norm of an $L^2$ normalized eigenstate goes
to zero in the limit $|\Lambda| \to \infty$.

What makes (\ref{1.5}) more difficult to analyse than (\ref{1.4})
is that the absolute value squared eliminates oscillations and
small denominators are felt when $\varepsilon$ is small. In fact,
even in finite volume (\ref{1.4}) does not diverge as
$\varepsilon$ goes to zero.  However, (\ref{1.5}) diverges roughly
like $\rho / (\varepsilon \xi^d)$ where $\xi$ is the length over
which the eigenfunctions are extended.  Roughly speaking, the
hyperbolic symmetry emerges because the two Green's functions in
(\ref{1.5}) have energies with imaginary parts of opposite signs.
Another feature of (\ref{1.5}) is that the corresponding
statistical mechanics model is not expected to decay rapidly in 3
dimensions but rather to exhibit a Goldstone mode so that
(\ref{1.5}) should behave in the limit $\varepsilon \to 0$ like
$1/|x-y|$, the Green's function of the Laplacian corresponding to
diffusive time evolution.

The main purpose of this article is to analyse Wegner's non-linear
sigma model (for one replica) as a component of Efetov's
supersymmetric model. More precisely we study a sigma model that
arises in Fyodorov's work \cite{fyodorov}.  This model is
formulated on a lattice, whereas Wegner's model emerges upon
taking a continuum limit.

Roughly speaking supersymmetric models of disordered quantum
systems have three sectors: the Boson-Boson, Fermion-Fermion and
Boson-Fermion sectors. The B-B sector has the hyperbolic symmetry;
this is the sector which we study. Although the field in this
sector may potentially diverge we show that for dimension $d \ge
3$, if $\varepsilon |\Lambda| \rho = 1$, all moments of the field
remain uniformly bounded.  This is the analogue (in the sigma
model approximation) of the conjectured bound on (\ref{1.5}) in $d
\ge 3$. The F-F sector may also be studied in the sigma model
approximation, and it corresponds to a classical Heisenberg model
taking values in the two-dimensional sphere. Considered on its
own, this sector has no divergence because of the compactness of
the target. In three dimensions, the nearest neighbor Heisenberg
model has an ordered state which may be established by using
infrared bounds. The main open problem which we do not address in
this article is the B-F sector, which couples the other two
sectors in a supersymmetric fashion. This must be understood to
obtain a complete picture of the SUSY models. Nevertheless, we
will see that many phenomena of interest are already reflected in
the sigma model analysed in this article.

We now describe the hyperbolic sigma model which we shall analyse.
In a periodic box $\Lambda \in \mathbb{Z}^d$ (not the $\Lambda$ of
before, but the lattice of cubes that tile the original lattice),
we associate to each lattice site $j \in \Lambda$ a matrix
\begin{equation}
  S_j = T_j^{\vphantom{-1}} \sigma_3^{\vphantom{\dagger}}
  T_j^{-1} \;,
\end{equation}
where $T_j$ is subject to the conditions $T^* \sigma_3 T =
\sigma_3 = \mathrm{diag}(1,-1)$ and $\mathrm{Det} \, T_j = 1$.
Thus $T_j$ belongs to $\mathrm{SU}(1,1)$, and $S_j$ belongs to an
adjoint orbit of $\mathrm{SU}(1,1)$, which may be identified with
the symmetric space $\mathrm{SU}(1,1) / \mathrm{U}(1) \cong
\mathrm{H}^2$ where $\mathrm{U} (1)$ is the isotropy subgroup
generated by $\mathrm{i} \sigma_3$.

The action or energy of a configuration $j \mapsto S_j$ is given by
\begin{equation}\label{1.7}
  A_{\Lambda} (S,h) = {\textstyle{\frac{1}{2}}}
  \sum\nolimits_\Lambda^\prime \mathrm{Tr}\, (S_j S_{j^\prime})
  + {\textstyle{\frac{1}{2}}} h \sum\nolimits_{j \in \Lambda}
  \mathrm{Tr}\, (\sigma_3 S_j) \;,
\end{equation}
where $\sum_\Lambda^\prime$ denotes the sum over pairs of
nearest-neighbor sites of $\Lambda$, and $h > 0$.  Let $d\mu(S)$
denote an invariant measure on $\mathrm{SU}(1,1) / \mathrm{U}(1)$
and define
\begin{displaymath}
  Z_{\Lambda}(\beta, h) = \int \mathrm{e}^{ - \beta \, A_{\Lambda}
  (S,h)} \prod\nolimits_{j \in \Lambda} d \mu(S_j) \;.
\end{displaymath}
Expectations in this model are given by
\begin{equation}\label{1.8}
  Z_{\Lambda}(\beta, h)^{-1} \int F(S)\, \mathrm{e}^{-\beta \,
  A_{\Lambda} (S,h)} \prod\nolimits_{j \in \Lambda} d \mu (S_j) =
  \langle F \rangle_{\Lambda} (\beta, h) \;.
\end{equation}
Note that by $M := S \sigma_3 = T T^\ast = M^\ast > 0$, the
$\mathrm {SU}(1,1)$-orbit $S = T \sigma_3 T^{-1}$ can be
identified with the positive Hermitian matrices $M$ in
$\mathrm{SU} (1,1)$.  Thus, $\mathrm{Tr}\, S_j S_{j^\prime} =
\mathrm{Tr}\, M_j^{\vphantom {-1}} M_{j^\prime}^{-1}$ and
$\mathrm{Tr}\, \sigma_3 S_j = \mathrm{Tr} \, M_j$, and we see that
$A_\Lambda(S,h) > 0$. Since an $\mathrm{SU} (1,1)$-symmetry
emerges at $h = 0$, positivity of $h$ is needed to make the
integrals exist.  $h$ corresponds to a magnetic field and breaks
the non-compact $\mathrm{SU}(1,1)$ symmetry to $\mathrm{U} (1)$.

Our main result may be stated as follows.
\begin{thm}\label{thm1}
For $d \geq 3$ space dimensions there is a constant $C_0$ such that
\begin{equation}
  \left\langle \big( \mathrm{Tr}\, \sigma_3 S_0 \big)^2
  \right\rangle_{\Lambda} (\beta,h) \leq C_0
\end{equation}
for all $\beta \geq 3/2$ and $|\Lambda| h \ge 1$.
\end{thm}
\begin{rem}
If the $\mathrm{SU}(1,1)$-symmetry were restored in the limit $h
\to 0$, the expectation of the unbounded observable $(\mathrm{Tr}
\, \sigma_3 S_0)^2$ would have to diverge in that limit. Our
result can therefore be viewed as a statement of spontaneous
symmetry breaking. (A more detailed discussion of what it means
for a non-compact symmetry to be broken spontaneously has recently
been given in \cite{ns}.)

Higher moments of $\mathrm{Tr}\, \sigma_3 S_0$ can also be bounded
in $d \ge 3$. For $d = 1, 2$ we expect (but do not prove) that the
same kind of bound holds except that we must require
\begin{equation}\label{1.10}
  |\Lambda| \leq \left\{ \begin{array}{ll} \exp (C_{2}\beta) \quad &d
    = 2 \;, \\ C_1 \beta \quad &d = 1 \;, \end{array} \right.
\end{equation}
with constants $C_1, C_2$ independent of $\beta$.

Our proof can easily be extended to finite-range interactions for
large $\beta$; however, for technical reasons it does not easily
extend to infinite-range interactions.
\end{rem}
In the next section we shall explain the relation between the band
random matrices and the sigma models described above.  Roughly
speaking, the magnetic field $h$ is proportional to $\varepsilon$,
the imaginary part of the energy in (\ref{1.5}), and $\beta$ is
proportional to $W^2$, the square of the band width.  In $d = 1$,
the action (\ref{1.7}) just describes a random walk on $\mathrm{H}
^2$ indexed by time $j \in \mathbb{Z}$.

In the sigma-model approximation we shall see that
\begin{eqnarray}
  \left\langle \left\langle \big| ( H - E + \mathrm{i}
  \varepsilon)^{-1}(x,x) \big|^2 \right\rangle \right \rangle
  \nonumber &\equiv& \frac{ \left\langle \big| ( H - E + \mathrm{i}
  \varepsilon)^{-1}(x,x) \big|^2 \, \big| \mathrm{Det}(H - E +
  \mathrm{i} \varepsilon) \big|^{-2} \right\rangle} { \left\langle
  \big| \mathrm{Det} ( H - E + \mathrm{i} \varepsilon ) \big|^{-2}
  \right\rangle} \nonumber \\ &\propto& \left\langle \big(
  \mathrm{Tr}\,\, \sigma_3 S_0 \big)^2 \right\rangle_{\Lambda}
  (\beta,h) \;. \label{1.11}
\end{eqnarray}
Here again we identify $h$ with $\epsilon$. The extra factors of the
inverse determinant appear because we have not included the F-F and
F-B sectors.

The proof of Theorem \ref{thm1} relies on the use of horospherical
coordinates $(s,t)$ to parametrize $S_j \in \mathrm{H}^2$ (for the
details see Sect.\ \ref{sect: coords}).  The action in these
coordinates is
\begin{equation}
  A_{\Lambda}(S,h) = \mathop{{\sum}'}\limits_{i,j \in \Lambda} \left(
  \cosh(t_i - t_j) + \mathrm{e}^{t_i + t_j}(s_i - s_j)^2 \right) + h
  \mathop\sum\limits_{j \in \Lambda} \left( \cosh t_j + s^2_j \mathrm
  {e}^{t_{j}} \right) \;.
\end{equation}
The Gibbs measure now takes the form $\mathrm{e}^{- \beta\, A_\Lambda
(S,h)} \prod_{j \in \Lambda} \mathrm{e}^{t_j} dt_j \, ds_j$.

Note that $A_\Lambda$ is convex in $t$ and quadratic in $s$.  (We
mention in passing that the Bakry-Emery tensor $\mathrm{Hess} +
\mathrm{Ricci}$ for $A_\Lambda$ is not positive in the natural
hyperbolic geometry.)  One of the key advantages of the
horospherical coordinates is that we can integrate out the $s$
variables thereby producing an effective action $E_h (t)$.  For
$\beta \geq 3/2$ we prove that the Hessian of $E_h (t)$ is
positive, and in fact as quadratic forms we prove
\begin{equation}
  E''(t) = \mathrm{Hess} \,\, E_h (t) \geq -(\beta -
  {\textstyle{\frac{1}{2}}} ) \Delta + h \;,
\end{equation}
with $\Delta$ the discrete Laplacian. Now the Brascamp-Lieb
inequality may be applied to control fluctuations of the $t_x$ in
terms of $\big( -(\beta - \frac{1}{2}) \Delta + h \big)^{-1}
(x,x)$. This is bounded in three dimensions provided $|\Lambda| h
\geq 1$.  In one or two dimensions one must require (\ref{1.10}).

The remainder of this paper is organized as follows.  In Sect.\
\ref{sect: origin} we describe the relation of the sigma model
(\ref{1.7}) to the random band model described by (\ref{1.3})
following ideas of Fyodorov.  Horospherical coordinates are
introduced in Sect.\ \ref{sect: coords} and convexity of the
effective action is proved in Sect.\ \ref{sect: int_s}.  The
Brascamp-Lieb inequality together with a Ward identity are used in
Sect.\ \ref{sect: BL_ineq} to obtain bounds on the $t$ fields. The
remaining Sects.\ \ref{sect: bounds} and \ref{sect: adjust}
explain how to control the $s$ field fluctuations and the
$h$-regularization.

There are a number of open questions related to this paper.  The
primary one is to determine whether there are analogous bounds for
more general hyperbolic sigma models such as those of higher rank.
There are also problems involving averages of Green's functions
which are not uniformly elliptic. Note that $s$ correlations are
expressed in terms of
\begin{equation}\label{1.14}
  \left\langle (- \nabla \mathrm{e}^{2t} \nabla + h \,
  \mathrm{e}^t)^{-1} \, (x,y) \right\rangle \;.
\end{equation}
Since the $t$ fields are not bounded from below, the Green's
function (\ref{1.14}) is not uniformly elliptic. The distribution
of the $t$ fields is given by the convex effective action $E_h$.
Although we obtain good bounds on (\ref{1.14}) for the diagonal $x
= y$, the off-diagonal bounds obtained by our methods are not
sharp.

\bigskip\noindent{\bf Acknowledgments:}
T.\ Spencer would like to thank M.\ Disertori, K.\ Gawedzki, G.\
Papanicolau and S.R.S.\ Varadhan for helpful comments.

\section{Origin of the model}\label{sect: origin}
\setcounter{equation}{0}

We now review how the non-linear sigma model, (\ref{1.7}) and
(\ref{1.8}), arises from the problem of computing Green's function
averages for some ensembles of band random matrices. Aside from
putting our analysis on a solid footing in random-matrix theory and
disordered electron physics, this review will explain the origin of
the hyperbolic target space $\mathrm{H}^2$ and its Riemannian
geometry.

Readers interested only in mathematical results, not in physical
motivation and background, are invited to skip the present section;
the remainder of the paper does not depend on it.

\subsection{Gaussian ensembles of band random matrices}

Let $\Lambda \subset \mathbb{Z}^d$ be a periodic box as before,
and assign to every site $i \in \Lambda$ one copy $V_i$ of an
$N$-dimensional unitary vector space.  (Physically speaking we are
assigning $N$ valence electron orbitals to every atom of a solid
with hypercubic lattice structure.) The finite-dimensional Hilbert
space $V$ of the random-matrix model to be specified is the
orthogonal sum
\begin{displaymath}
  V = V_1 \oplus V_2 \oplus \ldots \oplus V_{|\Lambda|} \;.
\end{displaymath}
The basic framework we have in mind is single-electron quantum
mechanics, and our goal is to establish control over the spectral
and transport properties of certain ensembles of random
Hamiltonians $H$. We shall take the Hamiltonians to be elements of
$\mathrm{Herm}(V)$, the space of Hermitian operators on $V$.

A random-matrix model is now defined by a probability distribution
on $\mathrm{Herm}(V)$.  Equivalently, one may specify the Fourier
transform or characteristic function:
\begin{equation}
  \Omega(K) = \big\langle {\rm e}^{ {\rm i} {\rm Tr}\, H K}
  \big\rangle \;,
\end{equation}
where $\langle \ldots \rangle$ denotes the expectation value
w.r.t.~the probability distribution for the random Hamiltonian
$H$. The Fourier variable is some other element $K \in \mathrm
{Herm}(V)$.

For simplicity we shall restrict ourselves to the case of Gaussian
distributions with zero mean, $\langle H \rangle = 0$.  If $\Pi_i$
is the orthogonal projector on the linear subspace $V_i \subset
V$, we take the characteristic function to be
\begin{equation}\label{2.2}
  \Omega(K) = \exp \left( - {\textstyle{\frac{1}{2}}} \sum\nolimits_{
    i, j = 1}^{|\Lambda|} J_{ij} \, {\rm Tr} \, (\Pi_i \, K \, \Pi_j
    \, K) \right) \;,
\end{equation}
where the coefficients $J_{ij}$ are real, symmetric, and
non-negative (they must also be positive semi-definite as a
quadratic form). The choice (\ref{2.2}) also implies that all
matrix entries of $H$ are statistically independent.

We mention in passing that the characteristic function (\ref{2.2})
is invariant under conjugation $K \mapsto U K U^{-1}$ by $U \in
\mathcal{U}$ where ${\mathcal U}$ is the direct product of all the
groups of unitary transformations in the subspaces:
\begin{displaymath}
  {\mathcal U} = {\rm U}(V_1) \times {\rm U}(V_2) \times \cdots
  \times {\rm U}(V_{|\Lambda|}) \;.
\end{displaymath}
This means that the probability distribution for the Hamiltonian
$H$ has a \emph{local gauge invariance}.  Models of this kind were
first introduced and studied by Wegner \cite{norbitals}.

By varying the lattice $\Lambda$, the number of orbitals $N$, and
the variances $J_{ij}$, one obtains a large class of Hermitian
random-matrix models.  In particular, if $d(i,j) = |i - j|$
denotes a distance function for $\Lambda$, and $f$ is a rapidly
decreasing positive function on $\mathbb{R}_+$ of width $W$, the
choice $J_{ij} = f(|i-j|)$ gives an ensemble of band random
matrices with band width $W$ and symmetry group $\mathcal{U} =
\mathrm{U}(N)^{|\Lambda|}$.

There exist two distinct situations where such a random-matrix
ensemble is expected to exhibit metallic behavior (in dimension $d
\ge 3$) and efficient methods of analysis are available.  The
first one occurs when the width $W$ is large (and, say, $N = 1$).
This case is dealt with by the Sch\"afer-Wegner method \cite{sw};
it will not be considered further in the present paper (see
however \cite{EMPsusymethod} for a recent review of that method).

The second one is called the `granular model'.  There, $N \gg 1$
and the diagonal of the variance matrix $J_{ij}$ dominates the
off-diagonal:
\begin{displaymath}
    J_{ii} \gg \sum_{j: j \not= i} J_{ij} \;.
\end{displaymath}
Each atom $i \in \Lambda$ here is to be viewed as a grain, or
small metallic particle, housing a large number $N$ of electron
states, and the squared matrix elements for tunneling between
grains ($J_{ij}$ for $i \not= j$) are small compared to the
intra-grain matrix elements $(J_{ii})$. The appropriate method to
use in this case is that of Fyodorov (Sect.\ \ref{sect:
fyodorov}). Metal\-lic behavior is expected to occur when the
coefficients $N J_{ij} / \sqrt{J_{ii} J_{jj}}$ are not too small.

Another (perhaps more familiar) way of defining the class of
granular models is to say that one starts from matrices $H$ drawn
from the Gaussian Unitary Ensemble (GUE) of matrix dimension $N
|\Lambda|$, partitions $V = \mathbb{C}^{N |\Lambda|}$ as $V = V_1
\oplus V_2 \oplus \ldots \oplus V_{|\Lambda|}$, and then
multiplies the variances of all matrix elements of $H$ connecting
$V_i \simeq \mathbb{C}^N$ with $V_j \simeq \mathbb{C}^N$ by the
positive number $J_{ij}$.  This is the same as the model
(\ref{1.3}) described in the introduction, with $N$ being equal to
the volume of the cubes tiling the lattice.

\subsection{Averaging the Green's functions over the disorder}

Fixing some lattice site $\ell \in \Lambda$, let $A_\ell$ be the
average absolute square of the $(\ell,\ell)$ part of the Green's
function with complex energy $E - \mathrm{i}\varepsilon$:
\begin{displaymath}
    A_\ell(E,\varepsilon) = \left\langle \big| {\rm Tr}\,
    (H - E + {\rm i} \varepsilon)^{-1} \Pi_\ell \big|^2
    \right\rangle \;.
\end{displaymath}
Physicists have developed an approximation scheme that calculates
disorder averages such as this one by relating them to the
correlation functions of a supersymmetric non-linear sigma model
\cite{efetov1}.  Here we shall address the related, but somewhat
simpler problem that arises from considering
\begin{equation}
    B_\ell^{(n)}(E,\varepsilon) = \frac{ \left\langle \big| {\rm Tr}\,
    (H - E + {\rm i} \varepsilon)^{-1} \Pi_\ell \big|^2 \times \big|
    {\rm Det} (H - E + \mathrm{i} \varepsilon) \big|^{-2n}
    \right\rangle } {\left\langle \big| \mathrm{Det} (H - E +
    \mathrm{i} \varepsilon) \big|^{-2n} \right\rangle} \;.
\end{equation}
Note that $B_\ell^{(n)}$ can be viewed as the average squared
Green's function for a \emph{deformed} ensemble, where the
probability distribution for the Hamiltonian $H$ is modified by
multiplying it with $2n$ inverse powers of $| {\rm Det}(H - E +
\mathrm{i}\varepsilon ) |$.  Inserting these factors modifies the
original problem, and it is far from clear how much bearing the
results for $n \ge 1$ will have on the case $n = 0$. (Physicists
often use a procedure called the replica trick, where one attempts
to infer the answer for $n = 0$ by analytic continuation from the
answer for $n \in \mathbb{N}$.  We will make no such attempt
here.) However, even if the $n \ge 1$ situation was quite
unrelated to $n = 0$, analysing it would still be a necessary step
toward establishing mathematical control over the full
supersymmetric theory at $n = 0$.  The reason is that Efetov's
supersymmetric non-linear sigma model has the effective theory of
$B_\ell^{(n)}$ at $n = 1$ for its non-compact bosonic sector.

In order to express $B^{(1)}_\ell$ we introduce a pair of complex
fields $\phi = (\phi_+, \phi_-)$ where $\phi_{\pm} \in V$.  The
projections $\Pi_j \, \phi_+ = \phi_+(j)$ and $\Pi_j \, \phi_- =
\phi_-(j)$ are complex $N$-component vectors. The scalar product is
given by
\begin{displaymath}
  (\phi_{+}^\ast, \phi_{+}^{\vphantom{\ast}}) = \sum\nolimits_j
  \phi_{+}^\ast (j) \cdot \phi_{+}^{\vphantom{\ast}} (j) \;.
\end{displaymath}
If $A$ is a matrix or linear operator acting on $V$ with $\mathfrak
{Re} \, A = \frac{1}{2} (A + A^*) > 0$, we normalize our Gaussian
integrals over $\phi$ such that
\begin{displaymath}
  \int \mathrm{e}^{- (\phi_{+}^{*}, \, A \,
  \phi_{+}^{\vphantom{*}})} = \mathrm{Det}^{-1} (A) \;.
\end{displaymath}
For fixed $E, \epsilon$ and $\Lambda$ define the quadratic form
\begin{equation}
  (\phi^*, \, A_H \phi) = - \mathrm{i} \big( \phi_+^*, (H - E +
  \mathrm{i} \varepsilon) \phi_+) + \mathrm{i} (\phi^*_- , (H - E -
  \mathrm{i} \varepsilon)\phi_- \big) \;.
\end{equation}
Note that $\mathfrak{Re}\, A \geq \varepsilon$ and the integral over
$\phi$ is therefore defined:
\begin{displaymath}
  Z_{\Lambda} \equiv \int \mathrm{e}^{- (\phi^{*}, \, A_{H} \phi)}
  = \big| \mathrm{Det} (H - E + \mathrm{i} \varepsilon) \big|^{-2} \;,
\end{displaymath}
and
\begin{equation}\label{obs 2.5}
  \int \mathrm{e}^{-(\phi^{*}, \, A_{H} \phi)} |\phi_+(\ell)|^2 \,
  |\phi_-(\ell)|^2 = \big| \mathrm{Tr}(H - E + \mathrm{i} \varepsilon)
  ^{-1} \Pi_\ell \big|^2 \, \big| \mathrm{Det}(H - E + \mathrm{i}
  \varepsilon) \big|^{-2} \;.
\end{equation}

For general $n \geq 1$ if we set $\phi_{\pm}(j) = \{ \phi_{\pm 1}(j),
\phi_{\pm 2}(j), \ldots, \phi_{\pm n}(j) \}$, this produces the factor
$| \mathrm{Det}(H - E + \mathrm{i} \varepsilon)|^{-2n}$ and permits us
to study more complicated observables involving several Green's
functions.

Now we can calculate the average of $Z_{\Lambda}$ over the
randomness in $H$ by using (\ref{2.2}).  First consider
\begin{equation}
  \left\langle \mathrm{e}^{ \mathrm{i} (\phi^{*}_{+}, \, H
    \phi_{+}^{\vphantom{\ast}}) - \mathrm{i} (\phi^{*}_{-}, \, H
    \phi_{-}^{\vphantom{\ast}}) } \right\rangle = \mathrm {e}^{-
    \frac{1}{2} \sum J_{ij} \mathrm{Tr}\, (s M_i \,s M_j )} \;,
\end{equation}
where
\begin{equation}\label{M 2.7}
  M_j = \begin{pmatrix} \phi_+^*(j) \cdot \phi_+(j) & \phi_+^*(j)
    \cdot \phi_{-}(j)\\ \phi^*_-(j) \cdot \phi_+(j) & \phi^*_-(j)
    \cdot \phi_-(j) \end{pmatrix} \;,
\end{equation}
and $s = \sigma_3 = \mathrm{diag}(1,-1)$. For general $n$, $\sigma_3$
is replaced by the diagonal matrix $s = \mathrm{diag} ( \mathrm{Id}_n,
- \mathrm{Id}_n)$.  Note that the $2n \times 2n$ matrices $M_j$ are
Hermitian and non-negative; we say they lie in $\mathrm{Herm}^+(
\mathbb{C}^{2n})$.

\subsection{Fyodorov's method}\label{sect: fyodorov}

Following Fyodorov \cite{fyodorov} we choose the matrices $M_j$ as
our integration variables, i.e., we push forward the measure over
the $\phi$ to a measure over the non-negative matrices $M$. The
push forward may be singular. However, if $N \geq 2n$ then the
push forward has a density (derived in Appendix A) given by
\begin{displaymath}
  \prod\nolimits_{i = 1}^{|\Lambda|} {\rm Det}^{N - 2n} (M_i) \, dM_i
  \;,
\end{displaymath}
where $dM_i$ denotes a (suitably normalized) Lebesgue measure on
$\mathrm{Herm}^+ (\mathbb {C}^{2n})$. Now set
\begin{equation}
    d\mu(M_i) := \mathrm{Det}^{-2n} (M_i) \, dM_i \;.
\end{equation}
Then we obtain Fyodorov's formula for $Z_\Lambda$ (for general $n$) in
the form
\begin{equation}\label{2.10}
    Z_\Lambda = \int {\rm e}^{- \frac{1}{2} \sum_{ij} J_{ij} \mathrm
    {Tr} \, ( s M_i\, s M_j )} \prod\nolimits_{k \in \Lambda} \,
    \mathrm{e}^{\mathrm {Tr}\, ( \mathrm{i} s E - \varepsilon)
      M_k}\, \mathrm{Det}^N (M_k) \, d\mu(M_k) \;,
\end{equation}
where the integral is over the configurations $\{ M_i \}$ with $M_i >
0$ for all $i \in \Lambda$.

The formulation (\ref{2.10}) offers a transparent view of the
symmetries of the problem.  Indeed, let $\mathrm{U}(n,n)$ be the
pseudo-unitary group of complex $2n \times 2n$ matrices $T$ with
inverse $T^{-1} = s T^\ast s$.  This group acts as a transformation
group on the matrices $M_i$ by
\begin{displaymath}
    M_i \mapsto T M_i T^\ast \;.
\end{displaymath}
Clearly, the integration domain for $M_i \in \mathrm{Herm}^+(
\mathbb{C}^{2n})$ of Fyodorov's formula (\ref{2.10}) is invariant
under that group action. Since $|{\rm Det}\, T| = 1$ for $T \in
\mathrm{U} (n,n)$, the same holds true for the integration density
$d\mu(M_i)$.  From
\begin{displaymath}
    M_i \, s \mapsto T M_i T^\ast s = T M_i \, s T^{-1}
\end{displaymath}
one sees that the function being integrated in (\ref{2.10})
becomes invariant under the $\mathrm{U}(n,n)$ group action when the
parameter $\epsilon$ is taken to zero.  Thus $\mathrm{U}(n,n)$
transformations are global symmetries in that limit.

In what follows we focus on the case $n = 1$, where the symmetry group
is $\mathrm{U}(1,1)$ or, what amounts to essentially the same,
$\mathrm{SU}(1,1)$.

\subsection{The sigma-model approximation}\label{sect: approx}

The exact integral representation (\ref{2.10}) is well suited for
further analysis in the granular limit which we now consider. Thus
we now assume $N \gg 1$, $J_{ij} = 0$ for $|i - j| \ge 2$, $J_{ij}
= J_1 > 0$ for $|i - j| = 1$, and
\begin{displaymath}
  J_{ii} \equiv J_0 \gg 2 J_1 d \;.
\end{displaymath}
Let us first consider the diagonal terms of the action
(\ref{2.10}):
\begin{displaymath}
  \sum\nolimits_{j \in \Lambda} \Big( {\textstyle{\frac{1}{2}}} J_0 \,
    \mathrm{Tr}\, (s M_j)^2 - \mathrm{i} E\, \mathrm{Tr}\, (s M_j ) -
    N \, \mathrm{Tr}\, \ln M_j \Big) \;.
\end{displaymath}
The matrices $M_j \, s$ may be expressed as
\begin{equation}
  M_j \, s = T_j \begin{pmatrix} p_1(j) &0 \\ 0 &- p_2(j) \\
  \end{pmatrix} T_j^{-1} \;,
\end{equation}
where $p_1(j)$, $p_2(j)$ are positive real numbers, and $T_j \in
\mathrm{SU}(1,1)$ is determined only up to right multiplication by an
arbitrary element in $K \equiv \mathrm{U}(1)$.  The measure becomes
\begin{equation}
  d\mu(M_j) = \frac{(p_1(j) + p_2(j))^2}{p_1(j)^2 p_2(j)^2} \,
  dp_1(j) dp_2(j) \, d\mu_K (T_j) \;,
\end{equation}
where $d\mu_K (T_j)$ is an invariant measure for $\mathrm{SU}(1,1) /
\mathrm{U}(1)$.  The diagonal terms of the action can be written in
terms of $p_1, p_2$:
\begin{displaymath}
  \frac{J_0}{2} (p_1^2 + p_2^2) - \mathrm{i} E (p_1 - p_2) -
  N( \ln p_1 + \ln p_2) \;.
\end{displaymath}
The critical point for $E^2 \le 4 N J_0$ is given by
\begin{equation}
  p_1 = \frac{\mathrm{i}E} {2 J_0} + \rho_N(E) \;, \quad
  \rho_N(E) = \frac{\sqrt{4 N J_0 - E^2}} {2 J_0} \;,
\end{equation}
and $p_2 = \bar{p}_1$, where $\rho_N$ is the mean-field expression for
the local density of states (but for a factor of $\pi$). In this
approximation,
\begin{equation}
  M_j \, s = T_j \begin{pmatrix} \rho_N(E) &0 \\ 0 &-\rho_N(E) \\
    \end{pmatrix} T_j^{-1} + \mathrm{const} \times \mathrm{Id} \;.
\end{equation}
The last term arising from the imaginary part of $p_1, p_2$ is of no
consequence.  Thus
\begin{equation}\label{saddle M_j}
  M_j \, s = \rho_N(E) S_j \;,
\end{equation}
where $S_j$ was defined by $S_j = T_j^{\vphantom{-1}}
\sigma_3^{\vphantom{\dagger}} T_j^{-1}$ as before.
The action function of our model now is
\begin{displaymath}
  A_\Lambda(S) = {\textstyle{\frac{1}{2}}} \beta
  \sum\nolimits_\Lambda^\prime \mathrm{Tr}\, (S_j S_{j'}) +
  {\textstyle{\frac{1}{2}}} h \sum\nolimits_{j \in \Lambda}
  \mathrm{Tr}\, (\sigma_3 S_j) \;,
\end{displaymath}
with $\beta = 2 J_1 \, \rho^2_N(E)$ and $h = 2 \varepsilon \,
\rho_N (E)$.  Similarly, by using (\ref{M 2.7}) and (\ref{saddle
M_j}), the observable appearing in (\ref{obs 2.5}) is proportional
to $(\mathrm{Tr}\, \sigma_3 S_0)^2$ in the sigma-model
approximation. The $\mathrm{SU}(1,1)$-invariant measure
$d\mu_K(T_j)$ is renamed to $d\mu(S_j)$.

In order to eliminate the sigma-model approximation we must
control the massive fluctuations of $p_1(j), p_2(j)$ about the
saddle. Although the Gibbs measure is complex, if we integrate
over these eigenvalues a new effective action is produced which is
real.  This new effective action may share the desired convexity
properties with $A_\Lambda(S)$.

\section{The model in horospherical coordinates}\label{sect: coords}
\setcounter{equation}{0}

Having clarified the origin of the hyperbolic non-linear sigma model
in disordered electron physics, we now begin our study of it.  In the
present section we introduce a coordinate system that takes advantage
of the hyper\-bolic structure of $\mathrm{H}^2$ and is well suited for
the purpose of doing analysis on the sigma model, Eqs.\ (\ref{1.7})
and (\ref{1.8}).

For any connected and simply connected noncompact Lie group $G$ with
semisimple Lie algebra there exists an Iwasawa decomposition
\cite{helgason}
\begin{displaymath}
  G = N A K \;,
\end{displaymath}
where $K$, $A$, and $N$ are maximal compact, maximal Abelian and
nilpotent subgroups, respectively. In the case at hand, namely $G
= \mathrm{SU}(1,1)$ with Lie algebra
\begin{displaymath}
  \mathfrak{su}(1,1) = \{ x_1 \sigma_1 + x_2 \sigma_2 + \mathrm{i} x_3
  \sigma_3 \big| (x_1,x_2,x_3) \in \mathbb{R}^3 \} \;,
\end{displaymath}
$K$ is the $\mathrm{U}(1)$ subgroup generated by $\mathrm{i}
\sigma_3$, and $ \sigma_i$ are the Pauli matrices.  We choose $A
\simeq \mathbb {R}^+$ to be the Abelian group generated by
$\sigma_1$; the nilpotent group $N$ then is the one-parameter
group with nilpotent generator $\sigma_2 - \mathrm{i} \sigma_3$.
Passing to equivalence classes or cosets by the right action of $K
= \mathrm{U}(1)$ on both sides of the Iwasawa decomposition, one
gets an identification
\begin{displaymath}
  \mathrm{H}^2 \simeq \mathrm{SU}(1,1) / \mathrm{U}(1) \simeq N A
  \cdot o \;.
\end{displaymath}
Thus the two-hyperboloid $\mathrm{H}^2$ is viewed as the orbit of
the one-parameter groups $N$ and $A$ acting on the coset $o = K$.

Introducing two real variables $s$ and $t$, we parameterize the
Lie groups $N$ and $A$ as
\begin{displaymath}
  N = \{ n_s = \mathrm{e}^{s(\sigma_2 - \mathrm{i}\sigma_3)/2} \big| s
  \in \mathbb{R} \} \;,\ \quad A = \{ a_t = \mathrm{e}^{t \sigma_1 /
  2} \big| t \in \mathbb{R} \} \;.
\end{displaymath}
We refer to $s$ and $t$ as \emph{horospherical} coordinates. Their
relation to the matrix $S$ para\-meterizing $\mathrm{H}^2$ is
given by
\begin{equation}
  S \, \sigma_3 = n_s a_t (n_s a_t)^\ast = \begin{pmatrix} \cosh t +
  \frac{s^2}{2} {\rm e}^t &\sinh t - ( \mathrm{i}s + \frac{s^2}{2})
  \mathrm{e}^t \\ \sinh t + ( \mathrm{i}s - \frac{s^2}{2}) \mathrm
  {e}^t &\cosh t + \frac{s^2}{2} {\rm e}^t \end{pmatrix} \;,
\end{equation}
and the $\mathrm{SU}(1,1)$-invariant metric tensor $g$ in these
coordinates takes the form
\begin{equation}
  g = {\textstyle{\frac{1}{2}}} \mathrm{Tr}\,\, \mathrm{d}S^2 =
  \mathrm{d}t^2 + \mathrm{e}^{2t} \mathrm{d}s^2 \;.
\end{equation}

How does the action of the subgroups $N$, $A$, and $K$ on
$\mathrm{H}^2$ look in horospherical coordinates?  (These group
actions are important because they furnish global symmetries of
the non-linear sigma model in the limit of vanishing
regularization, $h \to 0$.) First of all, since $N$ is a
one-parameter group one has
\begin{displaymath}
  n_{s_0} (n_s a_t) \cdot o = n_{s + s_0} a_t \cdot o \;,
\end{displaymath}
so $n_{s_0} \in N$ acts on $n_s a_t \cdot o \in NA \cdot o$ by
simply translating $(s,t) \mapsto (s + s_0, t)$. Second, from the
fact that $\sigma_2 - \mathrm{i} \sigma_3$ is an eigenvector of
the commutator action $[ \sigma_1 , \cdot ]$ with eigenvalue $-2$,
one easily verifies
\begin{displaymath}
  a_{t_0} (n_s a_t) \cdot o = (a_{t_0} n_s a_{-t_0}) (a_{t_0} a_t)
  \cdot o = n_{\mathrm{e}^{-t_0} s} \, a_{t + t_0} \cdot o \;,
\end{displaymath}
so $a_{t_0} \in A$ acts by $(s,t) \mapsto (\mathrm{e}^{-t_0} s , t
+ t_0)$. Third, the group action of $K$ in horospherical
coordinates is somewhat complicated and will not be considered
here.

The energy or action function of the non-linear sigma model
(\ref{1.8}) in horospherical coordinates is expressed by
\begin{equation}\label{3.3}
  A_\Lambda = \beta \sum\nolimits_{\Lambda}^\prime \left( \cosh(t_i -
  t_{i^\prime}) + {\textstyle{\frac{1}{2}}} (s_i - s_{i^\prime})^2
  \mathrm{e}^{t_i + t_{i^\prime}} \right) + h \sum\nolimits_{j \in
  \Lambda} \left( \cosh t_j + {\textstyle{\frac{1}{2}}} s^2_j \,
  \mathrm{e}^{t_j} \right) \;.
\end{equation}
The Gibbs measure is
\begin{displaymath}
  d\mu_{\Lambda,A} = \mathrm{e}^{-A_\Lambda} \prod\nolimits_{i
  \in \Lambda} \mathrm{e}^{t_i} dt_i \, ds_i \;.
\end{displaymath}
As expected, $d\mu_{\Lambda,A}$ becomes invariant under global $N$ and
$A$ transformations,
\begin{eqnarray}
  n_s : \quad (s_i\, , \, t_i) &\mapsto& (s_i + s \, , \, t_i) \;,
  \nonumber \\
  a_t : \quad (s_i\, , \, t_i) &\mapsto& (\mathrm{e}^{-t} s_i\, ,
  \, t_i + t) \label{A-symmetry} \;,
\end{eqnarray}
in the limit $h \to 0$.

Our observable given in Theorem \ref{thm1} may be expressed as
\begin{equation}\label{3.4}
  \big( \mathrm{Tr}\,\, \sigma_3 S_0 \big)^2 = \big( 2 \cosh \, t_0 +
  s_0^2 \, \mathrm{e}^{t_{0}} \big)^2 \;.
\end{equation}
Note that since the action is quadratic in $s$, the integral over
the variable $s_0$ is Gaussian and can be done explicitly.

\section{Integration of the $s$ fields}\label{sect: int_s}
\setcounter{equation}{0}

In this section we shall analyse the action (\ref{3.3}). Since it
is quadratic in the $s$ fields, they can be integrated out.
Consider the interaction between the $s$ and $t$ fields in
(\ref{3.3}) and define
\begin{equation}\label{cov s}
  B(s,t) = \sum\nolimits_{\Lambda}^\prime \mathrm{e}^{t_i + t_j} (s_i
  - s_j)^2 \equiv (s , D\, s)_\Lambda \;,
\end{equation}
where $(f , g)_{\Lambda} \equiv \sum_{i \in \Lambda} f_i\, g_i$,
and $D$ is a matrix corresponding to an elliptic operator with
periodic boundary conditions and with coefficients that depend on
$t_i$.  As a quadratic form $D$ is non-negative, and its matrix
elements are given by
\begin{equation}\label{defD}
  D_{ij} = \left\{ \begin{array}{l@{\quad}l} - \mathrm{e}^{t_i + t_j}
  &|i - j|=1 \\ 0 \qquad & |i - j| > 1 \end{array} \right. \;, \qquad
  D_{ii} = - \sum_{j: j \not= i} D_{ij} \;.
\end{equation}
When the variables $t_i$ all vanish, $D = -\Delta_{\Lambda}$ where
$\Delta_{\Lambda}$ is the discrete Laplacian of the lattice $\Lambda$
with periodic boundary conditions. Although $D$ is elliptic, it is not
uniformly elliptic as the $|t_i|$ may be very large.

Using (\ref{cov s}) and integrating over the $s$ fields we obtain
an explicit expression for the effective action:
\begin{equation}
  E_h = \beta \sum\nolimits_{\Lambda}^\prime \cosh \ (t_i - t_{i
  ^\prime}) + C_h(t) + \sum\nolimits_{j \in \Lambda} ( - t_j + h \,
  \cosh t_j ) \;,
\end{equation}
where
\begin{equation}\label{4.4}
  C_h(t) = {\textstyle{\frac{1}{2}}} \ln \mathrm{Det} \left( D(t) +
  h \, \mathrm{e}^t \right) + \mathrm{const} = - \ln \int \mathrm{e}
  ^{-\frac{\beta}{2} B(s,t) - \frac{h}{2} \sum \mathrm{e}^t s^2}
  {\textstyle{\prod_{i \in \Lambda}}}\, ds_i \;.
\end{equation}
We are going to regard the torus variables $t_i$ as Cartesian
coordinates of $\mathbb{R}^{|\Lambda|}$ equipped with the canonical
Euclidean geometry, and have therefore relocated the variable volume
factors $\mathrm{e}^{t_i}$ from $\prod \mathrm{ e}^{t_i} dt_i$ to
$E_h$. Notice that the effective Gibbs measure
\begin{displaymath}
  \mathrm{e}^{-E_h} \, {\textstyle{\prod_{i \in \Lambda}}} \, dt_i
\end{displaymath}
for $h = 0$ is invariant under shifts $t_i \to t_i + \gamma$. This
invariance is a remnant of the global symmetry (\ref{A-symmetry}) of
the original theory, and will play an important role in later
discussions.

We shall first analyse a slightly different expression
\begin{equation}\label{4.5}
  \mathrm{e}^{- \widetilde{C}(t)} = \int \mathrm{e}^{- \frac{\beta}
  {2} B(s,t)} \delta ({\textstyle{\sum_{i \in \Lambda}}}\, s_i ) \,
  {\textstyle{\prod\nolimits_{j \in \Lambda}}}\, ds_j \;,
\end{equation}
where the $\delta$-function eliminates the zero mode of $B$ and
makes the integral exist.  By Gaussian integration we have
\begin{displaymath}
  \widetilde{C}(t) = {\textstyle{\frac{1}{2}}} \ln \mathrm{Det} \,
  \widetilde{D}(t) \;,
\end{displaymath}
where $\widetilde{D} > 0$ is $D$ acting on the orthogonal
complement of the constant functions.  Both $\widetilde{C}$ and
$\widetilde{D}$ depend on $\Lambda$ and the $t$ field, but we
shall frequently omit these dependences for notational brevity.
The effective action in $t$ is
\begin{equation}\label{4.6}
  E_\Lambda = \beta \sum\nolimits_{\Lambda}^\prime \cosh(t_i -
  t_{i^\prime}) + \widetilde{C}(t) + \sum\nolimits_{j \in \Lambda}
  (- t_j + h\, \cosh t_j) \;.
\end{equation}
We shall set the factor of $\beta /2$ appearing in (\ref{4.5})
equal to 1. By scaling in $s$ this simply shifts $\widetilde{C}
(t)$ by a trivial constant.

For a function $F$ of $t_i$ ($i \in \Lambda$) let the Euclidean
Hessian of $F$ be denoted by $F''$:
\begin{displaymath}
  F_{ij}^{\prime\prime} = \frac{\partial^2 F}{\partial t_i
  \partial t_j} \quad (i,j \in \Lambda) \;.
\end{displaymath}
\begin{thm}\label{thm2} For any value of the coupling parameter
$\beta \geq 3/2$ and dimension $d \geq 1$ the function $E_{
\Lambda}$ is convex and
\begin{displaymath}
  E_{\Lambda}^{\prime\prime} \geq - (\beta - {\textstyle{\frac{1}
  {2}}}) \Delta_\Lambda + h \geq - \Delta_\Lambda + h \;.
\end{displaymath}
\end{thm}
\medskip\noindent\emph{Proof}. --- Clearly from (\ref{4.6})
\begin{equation}\label{obv-est}
  E_\Lambda^{\prime\prime} \geq -\beta \, \Delta_\Lambda + h +
  \widetilde{C}'' \;,
\end{equation}
so it suffices to estimate $\widetilde{C}''$.  From (\ref{4.5})
with $\beta / 2$ set to 1 we have
\begin{equation}\label{defU}
  \frac{\partial \widetilde{C}}{\partial t_i} = \langle U_i
  \rangle_{s^{\vphantom{'}}} \;, \qquad U_i = \sum\limits_{j :
  |i - j| = 1} \mathrm{e}^{t_i + t_j} (s_i - s_j)^2 \;,
\end{equation}
where $\langle \cdot \rangle_{s^{\vphantom{'}}}$ denotes the
average over the $s$ field with Gibbs weight $\mathrm{e}^{-B}
\delta$. For $|i - j| > 1$ we have
\begin{displaymath}
  - \frac{\partial^2 \widetilde{C}}{\partial t_i \partial t_j} =
  \langle U_i\, ; \, U_j \rangle_{s^{\vphantom{'}}} \equiv \langle
  U_i \, U_j \rangle_{s^{\vphantom{'}}} - \langle U_i \rangle_{
  s^{\vphantom{'}}} \langle U_j \rangle_{s^{\vphantom{'}}} \;,
\end{displaymath}
and for $|i - j| = 1$
\begin{displaymath}
  - \frac{\partial^2 \widetilde{C}}{\partial t_i \partial t_j} =
  \langle U_i \, ; \, U_j \rangle_{s^{\vphantom{'}}} - \langle
  \mathrm{e}^{t_i + t_j} (s_i - s_j)^2 \rangle_{s^{\vphantom{'}}} \;,
\end{displaymath}
while on the diagonal
\begin{displaymath}
  - \frac{\partial^2 \widetilde{C}}{\partial t_i^2} = \langle U_i \, ;
  \, U_i \rangle_{s^{\vphantom{'}}} - \langle U_i \rangle_{s^{
  \vphantom{'}}} \;.
\end{displaymath}
Let
\begin{equation}\label{defK}
  K_{ij} = \langle U_i \, ; \, U_j \rangle_{s^{\vphantom{'}}} \;,
\end{equation}
and decompose $\widetilde{C}^{\prime\prime}$ into two pieces:
\begin{equation}\label{decomp-C}
  \widetilde{C}_{ij}^{\prime\prime} = \left( 2 \langle U_i
  \rangle_{s^{\vphantom{'}}} \delta_{ij} - K_{ij} \right) + R_{ij} \;,
\end{equation}
where $R$ is a local remainder term.

By explicit computation $K_{ij}(t) \geq 0$ for all $i, j$ and field
configurations $t$, because the square of a Green's function arises.
\begin{lem} With $U$ and $K$ defined by (\ref{defU}) and
(\ref{defK}) we have for every $t$
\begin{equation}\label{4.11}
  \sum\nolimits_j K_{ij} = 2 \langle U_i \rangle_{s^{\vphantom{'}}}
  \;.
\end{equation}
This relation implies that as a quadratic form
\begin{equation}\label{4.12}
  2 \langle U_i \rangle_{s^{\vphantom{'}}} \, \delta_{ij} - K_{ij}
  \geq 0 \;.
\end{equation}
\end{lem}
\begin{proof} To get (\ref{4.11}) make the change of variables
$s_k \to s_k \mathrm{e}^\gamma$ ($k \in \Lambda$).  Then $U_i \to U_i
\, \mathrm{e}^{2 \gamma}$ and $B \to B \, \mathrm{e}^{2 \gamma}$,
while the expectation value $\langle U_i \rangle_{s^{ \vphantom{'}}}$
remains invariant. Differentiating $\langle U_i \rangle_{s^{
\vphantom{'}}}$ with respect to $\gamma$ at $\gamma = 0$ yields
\begin{displaymath}
  0 = 2 \langle U_i \rangle_{s^{\vphantom{'}}} - 2 \langle U_i; B
  \rangle \;,
\end{displaymath}
and since $B = \frac{1}{2} \sum_j U_j$ we obtain (\ref{4.11}).

The non-negativity of the quadratic form (\ref{4.12}) now follows from
the Schwarz inequality:
\begin{displaymath}
  \Big| \sum\limits_{i,j} K_{ij} f_i f_j \Big| \leq \Big[
  \sum\limits_{i,j} K_{ij}^{\vphantom{2}} f_j^2 \Big]^{\frac{1}{2}} \,
  \Big[ \sum\limits_{i,j} K_{ij}^{\vphantom{2}} f_i^2
  \Big]^{\frac{1}{2}} = 2 \sum\limits_i U_i^{\vphantom{2}} f_i^2 \;.
\end{displaymath}
Here we used the pointwise positivity of $K_{ij}$ to write the first
expression as a scalar product of two vectors $u$ and $v$ with
components $u_{ij} = \sqrt{K_{ij}} f_j$ and $v_{ij} = \sqrt{K_{ij}}
f_i$.
\end{proof}
We now must estimate the remaining local part $R = \widetilde{C}^{
\prime\prime} - 2 \langle U \rangle \delta + K$, which is expressed by
\begin{displaymath}
  R_{ij} = - \langle U_i \rangle_{s^{\vphantom{'}}} \delta_{ij} +
  \mathrm{e}^{t_i + t_j} \langle (s_i - s_j)^2 \rangle_{s^{
  \vphantom{'}}} \delta(|i-j| - 1) \;.
\end{displaymath}
Note that $R \leq 0$ as a quadratic form and that $\sum_j R_{ij} = 0$
for each $i$.
\begin{lem} For all real $f_i$
\begin{equation}\label{4.13}
  \Big| \sum\limits_{i,j} R_{ij} f_i f_j \Big| \leq {\textstyle{
  \frac{1}{2}}} \sum\limits_i (\nabla f)_i^2 \;,
\end{equation}
where $\nabla f$ denotes the discrete gradient of the lattice
$\Lambda$.
\end{lem}
\begin{proof} The left-hand side of (\ref{4.13}) can be written as
a sum over nearest-neighbor pairs $i,j$:
\begin{displaymath}
  \sum\nolimits_{|i-j|=1}^\prime R_{ij} (f_i - f_j)^2 \;.
\end{displaymath}
It therefore suffices to show that for each pair $i,j$ we have
\begin{displaymath}
  \mathrm{e}^{t_i + t_j} \langle (s_i - s_j)^2 \rangle_{
  s^{\vphantom{'}}} \leq 1/2 \;.
\end{displaymath}
This result follows from the fact that $\langle \, \cdot \, \rangle_s$
is a Gaussian expectation in $s$ with terms $\mathrm{e}^{ t_i + t_j}
(s_i - s_j)^2$ appearing in the action, $B$. Indeed, if $u, v_1, v_2,
\ldots$ are real variables it is a general fact that
\begin{displaymath}
  \frac{\int cu^2 \, \mathrm{e}^{- cu^2 - Q(u,v)} du\, \prod_a dv_a}
  {\int \mathrm{e}^{- cu^2 - Q(u,v)} du\, \prod_a dv_a} \leq 1/2
\end{displaymath}
for any positive constant $c$ and any $Q \geq 0$ which is quadratic in
$u,v$.  If we set $cu^2 = (s_i - s_j)^2 \mathrm{e} ^{t_i + t_j}$ and
$Q = B - cu^2$ (restricted to the linear subspace given by the
constraint $\sum_{i \in \Lambda} s_i = 0$), we obtain the desired
result.
\end{proof}
{}From the decomposition (\ref{decomp-C}) and the two lemmas we have
\begin{displaymath}
  {\widetilde{C}}'' \ge R \ge {\textstyle{\frac{1}{2}}} \Delta_\Lambda
  \;.
\end{displaymath}
Inserting this bound into (\ref{obv-est}) completes the proof of
Theorem \ref{thm2}.

\section{The Brascamp-Lieb Inequality}\label{sect: BL_ineq}
\setcounter{equation}{0}

We now state the Brascamp-Lieb inequality \cite{BL} in a form in which
we shall apply it.  Let $A = A(t)$ be a convex function of $N$
variables $t = (t_1, \ldots, t_N) \in \mathbb{R}^N$, where $\mathbb{R}
^N$ is the Euclidean vector space with scalar product
\begin{displaymath}
   (\varphi,t) = \sum\nolimits_{i = 1}^N \varphi_i \, t_i \;.
\end{displaymath}
With the function $A$ associate the measure $d\mu_A(t) =
\mathrm{e}^{-A(t)} \prod dt_i$.  Assume that the Euclidean
Hessian of $A$ satisfies
\begin{equation}\label{5.1}
    A''(t) \geq H > 0 \;,
\end{equation}
where $H$ is a positive $N \times N$ matrix independent of $t$.
\begin{thm}[Brascamp-Lieb] If $A$ satisfies (\ref{5.1}) then
\begin{equation}
  \langle \mathrm{e}^{(\varphi,\, t)} \rangle_A = \frac{\int
  \mathrm{e}^{(\varphi,\, t)} \mathrm{e}^{- A(t)}\prod^N dt_i}{\int
  \mathrm{e}^{- A(t)}\prod^N dt_i} \leq \mathrm{e}^ {\langle (\varphi
  ,\, t) \rangle_A} \, \mathrm{e}^{ \frac{1}{2}(\varphi, H^{-1}
  \varphi)} \;.
\end{equation}
\end{thm}
For our application we identify $A$ with the function $E =
E_\Lambda (t)$ given by (\ref{4.6}).  Theorem \ref{thm2} tells us
to put $H = - (\beta - \frac{1}{2}) \Delta_\Lambda + h$.  If we
then set
\begin{equation}\label{5.3}
    G_{ij} = {\left( -(\beta - {\textstyle{\frac{1}{2}}})
    \Delta_{\Lambda} + h \right)^{-1}}_{ij}
\end{equation}
and fix a site $i \in \Lambda$ with field variable $t_i$, we have
\begin{equation}\label{5.4}
  \langle \mathrm{e}^{ \alpha t_i - \alpha \langle t_i \rangle_E}
  \rangle_E \leq \mathrm{e}^{\frac{1}{2} \alpha^2 G_{ii}} \;.
\end{equation}
In dimension $d \geq 3$ and for $\beta \ge 3/2$ the Green's function
$G_{ii}$ is uniformly bounded as $\Lambda \to \mathbb{Z}^d$ provided
$h |\Lambda| \geq 1$.

Now we drop the subscript $E$ and let $\langle \cdot \rangle = \langle
\cdot \rangle_E$.  Large fluctuations of the field $t$ away from its
average are very unlikely:
\begin{displaymath}
  p = \mathrm{Prob}_E \{ t_i - \langle t_i \rangle \geq \rho \} \leq
  \mathrm{e}^{-\frac{\rho^2}{2G_{ii}}} \;.
\end{displaymath}
Indeed, when $t_i - \langle t_i \rangle \geq \rho$ we have
$\mathrm{e}^{ \alpha \,( t_i - \langle t_i \rangle - \rho )} \geq
1$ for all $\alpha \geq 0$, so that
\begin{displaymath}
  p \le \langle \mathrm{e}^{ \alpha \, (t_i - \langle t_i \rangle -
  \rho) } \rangle \leq \mathrm{e}^{- \alpha\, \rho} \, \mathrm{e}^
  {\frac{1}{2} \alpha^2 G_{ii}} = \mathrm{e}^{ -\frac{\rho^2}{2
  G_{ii}}}
\end{displaymath}
for $\alpha = \rho / G_{ii}$.  The same estimate applies to the
probability of an event $t_i - \langle t_i \rangle \le -\rho$, so
altogether we have
\begin{equation}\label{unlikely}
  \mathrm{Prob}_E \{ | t_i - \langle t_i \rangle | \geq \rho \} \leq
  2\, \mathrm{e}^{-\frac{\rho^2}{2G_{ii}}} \;.
\end{equation}

Our estimates on the $t$ field will be complete once we have estimated
the average $\langle t_i \rangle$. To do this consider the change of
variables $t_j \to t_j + \gamma$ , $s_j \to \mathrm{e}^{- \gamma}\,
s_j$ $(j \in \Lambda)$. Then if we take the derivative in $\gamma$ of
the logarithm of the partition function at $\gamma = 0$ we get
\begin{equation}
  h \sum\nolimits_{j \in \Lambda} \langle \sinh t_j \rangle = 1 \;,
\end{equation}
where 1 is produced from the $\delta$-function: $\delta
(\mathrm{e} ^{-\gamma} \sum s_j) = \mathrm{e}^{\gamma}\,
\delta(\sum s_j)$.  By translation invariance we see that for $h
|\Lambda| = 1$ we have $\langle \sinh t_i \rangle = 1$, so that by
Jensen's inequality and (\ref{5.4}) we have
\begin{displaymath}
  \mathrm{e}^{\langle t_i \rangle} \leq \langle \mathrm{e}^{t_i}
  \rangle = 2 + \langle \mathrm{e}^{- t_i} \rangle \leq 2 +
  \mathrm{e}^{- \langle t_i \rangle} \, \mathrm{e}^{\frac{1}{2}
  G_{ii}} \;.
\end{displaymath}
This gives an upper bound to $\langle t_i \rangle$:
\begin{displaymath}
  \langle t_i \rangle \leq 1 + {\textstyle{\frac{1}{4}}} G_{ii} \;.
\end{displaymath}
To obtain the lower bound we use $\langle \sinh t_i \rangle \geq
0$:
\begin{displaymath}
  \mathrm{e}^{-\langle t_i \rangle} \leq \langle \mathrm{e}^{-t_i}
  \rangle \leq \langle \mathrm{e}^{t_i} \rangle \leq \mathrm{e}^{
  \langle t_i \rangle} \, \mathrm{e}^{\frac{1}{2} G_{ii}} \;.
\end{displaymath}
Hence
\begin{equation}\label{lu-bound}
  - {\textstyle{\frac{1}{4}}} G_{ii} \leq \langle t_i \rangle \leq 1 +
  {\textstyle{\frac{1}{4}}} G_{ii} \;.
\end{equation}
This completes our estimates on the $t$ fields and its fluctuations.

\section{Bounds on the $s$ fields.}\label{sect: bounds}
\setcounter{equation}{0}

Recall that in addition to the $t$ variables the observable given
by (\ref{3.4}) contains factors of $s_0^2$ and $s_0^4$.  These
averages may be explicitly calculated in terms of the covariance
for the $s$ field given, see (\ref{cov s}), by $D^{-1}$ on the
orthogonal complement of the constant functions which we have
denoted by $\widetilde D^{-1}$.  Here we show how to deal with the
$s_0^4$ term:
\begin{equation}
  \langle s_0^4 \rangle_s^{\vphantom{\dagger}} = 3 \widetilde D^{-1}
  (0,0)^2 \;.
\end{equation}
The $s_0^2$ term is similar and can be handled in the same way.

The operator $D$ is non-negative but depends on $t$.  If all $t_j \geq
0$ then $D \geq -\Delta_{\Lambda}$ and we have bounds on $D^{-1}$ in
terms of the free Green's function $-\Delta^{-1}_{\Lambda}$ in
dimension three. However, the $t$ field may take large negative values
and so there is no uniform bound on $\widetilde{D}^{-1}(t)$.  The
control of $\langle \widetilde{D}^{-1}(0,0)^2 \rangle$ will come
from the fact that large negative values of $t$ are very rare by
(\ref{unlikely}) and (\ref{lu-bound}).

To bound the average of $\widetilde D^{-1}(0,0)^2$ we shall first
consider an elliptic operator $L$ whose quadratic form is
\begin{displaymath}
  (f, L f) \equiv \sum\nolimits_{j \in \Lambda} (\nabla f)^2_j \,
  a_j^{\vphantom{2}} \;,
\end{displaymath}
where $a_j \ge (1+|j|)^{-p}$. Let $\widetilde{L}$ denote the
projection on the orthogonal complement of the constant functions.
\begin{lem}\label{lem3} For $d \geq 3$ and $p < d-2$ the Green's
function of $\widetilde{L}$ is uniformly bounded (as $\Lambda \to
\mathbb{Z}^d$) by $0 \leq \widetilde {L}^{-1} (0,0) \leq A_p <
\infty$.
\end{lem}
\begin{proof}
Let $C_n$ denote the cube of side $2^n$ centered at the origin and
let $\chi_n$ be its indicator function. Note that
\begin{displaymath}
  f_n = 2^{-dn} \chi_n - 2^{-d(n+1)}\chi_{n+1}
\end{displaymath}
has zero average and the square of its $L^2$ norm is bounded by
$2^{-d n}$.

As a quadratic form, $\widetilde L$ restricted to $C_{n}$ (with
Neumann boundary conditions) is at least $2^{-(2+p)n}$, and we
therefore have
\begin{displaymath}
  (f_{n-1} \, , \widetilde{L}^{-1} f_{n-1})
  \leq 2^{(2+p)n} 2^{-d(n-1)} \;.
\end{displaymath}
To complete the proof of the lemma note that the projection of
$\delta_0$ onto the orthogonal complement of the constants can be
written as a sum over the $f_n$. By the Schwarz inequality we have
\begin{displaymath}
  \widetilde{L}^{-1}(0,0) \leq \left( \sum\nolimits_{n=0}^\infty
  ( f_n, \widetilde{L}^{-1} f_n )^{1/2} \right)^2 \leq A_p \;,
\end{displaymath}
provided that $p < d - 2$.
\end{proof}
\begin{lem} There is a constant $c_0$ so that
\begin{equation}\label{6.4}
  \left\langle \widetilde{D}^{-1}(0,0)^2 \right \rangle \leq c_0 \;.
\end{equation}
\end{lem}
\begin{proof}
Fix some value of $p$ with $0 < p \le \frac{1}{2}$, and for each
integer $k$ let $\chi_k(t)$ denote the characteristic function of
the set of configurations $t = \{ t_j \}_{j \in \Lambda}$ that
satisfy
\begin{displaymath}
\mathrm{e}^{t_x + t_y} \geq \mathrm{e}^{-k} (|x|+1)^{-p}
\end{displaymath}
for all nearest neighbors $x, y$.  We then have $\chi_k(t)
\widetilde{D}^{-1}(0,0)^2 \leq (A\, \mathrm{e}^k)^2$ by Lemma
\ref{lem3}.

We now claim that for all $k \ge \kappa \equiv \mathrm{max} \{ 0,
- \langle t_x + t_y \rangle \}$,
\begin{equation}\label{6.5}
  \left\langle 1 - \chi_k(t) \right\rangle_t \leq
  B\, \mathrm{e}^{-c \, (k - \kappa)^{2}} \;,
\end{equation}
which by Borel-Cantelli gives the desired statement:
\begin{eqnarray*}
  \left\langle \widetilde{D}^{-1}(0,0)^2 \right\rangle &=&
  \left\langle \big( \chi_0 + \chi_1 (1 - \chi_0) + \chi_2 (1 -
  \chi_1) + \ldots \big) \, \widetilde {D}^{-1}(0,0)^2 \right\rangle
  \\ &\leq& \mathrm{const} + A^2 B \sum\limits_{k \ge \kappa}
  \mathrm{e}^{2(k+1)} \mathrm{e}^{- c \, (k - \kappa)^2} \leq
  \mathrm {const} \;.
\end{eqnarray*}
To establish our claim (\ref{6.5}) suppose that
\begin{displaymath}
  \mathrm{e}^{t_{x}+t_{y}} \leq \frac{\mathrm{e}^{-k}}{(|x|+1)^p} \leq
  \mathrm{e}^{-(k+np)}
\end{displaymath}
for $x$ in the range $\mathrm{e}^n \leq |x| \leq \mathrm{e}^{n +
1}$. Then by (\ref{unlikely}) and (\ref{lu-bound}) the probability
of this event for $k \ge \kappa$ is less than
\begin{displaymath}
  \mathrm{e}^{(n+1) d} \, \mathrm{e}^{-c (k - \kappa + np)^2} \;,
\end{displaymath}
\noindent whose sum over $n$ is no greater than $\mathrm{e}^{-c\,
(k - \kappa)^{2}}$ times a constant $B$.
\end{proof}
The lemma works for any power of $\widetilde{D}^{-1}(0,0)$. Thus
by the Schwarz inequality and (\ref{5.4}) we can bound the
expectation of our observable $\mathrm{e}^{2 t_0} s_0^4$.

\section{Adjusting the regularization}\label{sect: adjust}
\setcounter{equation}{0}

We have used the $\delta$-function regularization in the $s$
variables rather than the correct term $h \sum_{j \in \Lambda}
s_j^2 \, \mathrm {e}^{t_j}$ which appears in the action
$A_\Lambda$.

Recall $C_h = \frac{1}{2} \ln \mathrm{Det}\, (D + h\, \mathrm{ e}^t)$
and $\widetilde{C} = \frac{1}{2} \ln \mathrm{Det}\, \widetilde D$. We
shall express $C_h$ in terms of $\widetilde{C}$.  To do this, let
$P_0$ denote the orthogonal projector on the vector space spanned by
the normalized constant function $\psi_{0,j} = |\Lambda|^{-1/2}$, and
let $P = 1 - P_0$.  The determinant can be calculated in terms of $P$
and $P_0$ blocks:
\begin{displaymath}
    \mathrm{Det}(D + h\, \mathrm{e}^t) = \mathrm{Det}
    (\widetilde{D} + h P_t)\, (\psi_0 , h\, \mathrm{e}^t \psi_0)
    \;,
\end{displaymath}
where $P_t$ is given by
\begin{displaymath}
    P_t = P\, \mathrm{e}^t P - P\, \mathrm{e}^t P_0
    \, \mathrm{e}^t P \cdot (\psi_0 , \mathrm{e}^t
    \psi_0)^{-1}.
\end{displaymath}
Using the Schwarz inequality it is easy to see that $P_t \geq 0$.

Thus $C_h = \widetilde{C} + \frac{1}{2} \mathrm{Tr} \ln (1 + h
\widetilde{D}^{ -1} P_t) + \frac{1}{2} \ln \ (\psi_0 , \mathrm{
e}^t \psi_0)$. We have left out the $\ln h$ term since it is
cancelled in the normalization.

Let $F(t)$ be our (positive) observable.  Now we can write
\begin{displaymath}
    \langle F(t)\, \rangle_{E_h} = \frac{\langle F(t)\,
    \mathrm{e}^{-R_h} \rangle_E}{ \langle \mathrm{e}^{-R_h}
    \rangle_E } \;,
\end{displaymath}
where
\begin{displaymath}
    R_h = C_h - \widetilde{C} =
    {\textstyle{\frac{1}{2}}} \mathrm{Tr}\, \ln (1 + h
    \widetilde{D}^{-1} P_t) + {\textstyle{\frac{1}{2}}}
    \ln \, (\psi_0 , \mathrm{e}^t \psi_0) \;.
\end{displaymath}
Since the first term of $R_h$ is positive we have
\begin{displaymath}
    \langle F(t) \rangle_{E_h} \leq \frac{\langle F(t)\, (\psi_0 ,
    \mathrm{e}^t \psi_0)^{- \frac{1}{2}} \rangle_E} {\mathrm{e}^{-
    \langle R_h \rangle_E}} \leq \langle F(t) \mathrm{e}^{- \frac{1}
    {2}(\psi_0, t\psi_0)} \rangle_E \,\, \mathrm{e}^{\langle R_h
    \rangle_E}
\end{displaymath}
where we have used Jensen's inequality. Since $P_t$ is positive we have
\begin{displaymath}
    \big\langle \mathrm{Tr}\, \ln ( 1 + h \widetilde{D}^{-1} P_t)
    \big\rangle_E \leq h\, \big\langle \mathrm{Tr}\, (\widetilde
    {D}^{-1} P_t) \big\rangle_E \leq h |\Lambda| \langle \mathrm
    {e}^{t_0} \widetilde{D}^{-1}(0,0)\rangle_E \leq \mathrm{const}
    \;,
\end{displaymath}
and the other term in $\langle R_h \rangle$ is estimated by
\begin{displaymath}
    \big\langle {\textstyle{\frac{1}{2}}}
    \ln \, (\psi_0 , \mathrm{e}^t \psi_0) \big\rangle \le
    {\textstyle{\frac{1}{2}}} \big\langle \mathrm{e}^{t_0}
    \big\rangle \le \mathrm{const} \;.
\end{displaymath}
The desired bound on $\langle F \rangle_{E_h}$ now follows from
estimates we obtained for $\langle \,\, \rangle_E$. This completes
our proof of Theorem \ref{thm1}.

\section{Appendix: Push forward of measure in Fyodorov's method}
\label{sect: appendix}
\setcounter{equation}{0}

Consider the mapping
\begin{displaymath}
  \psi : \mathrm{Hom}(\mathbb{C}^n , \mathbb{C}^N) \to \mathrm{Herm}^+
    (\mathbb{C}^n) \;, \quad \varphi \mapsto \varphi^\ast \varphi = M
    \;,
\end{displaymath}
and fix some (translation-invariant) Lebesgue measure $d\varphi \,
d\bar\varphi$ on $\mathrm{Hom}(\mathbb{C}^n , \mathbb{C}^N)$. We claim
that, if $N \ge n$, there exists a Lebesgue measure $c_{n,N}\, dM =
dM_{n,N}$ (with normalization constant depending on $n$ and $N$) such
that the equality
\begin{equation}\label{bosonize}
  \int\limits_{\mathrm{Hom}(\mathbb{C}^n, \mathbb{C}^N)} F( \varphi
  ^\ast \varphi ) \, d\varphi \, d\bar\varphi \, = \int\limits_
  {\mathrm{Herm}^+(\mathbb{C}^n)} F(M) \, \mathrm{Det}^{N-n} (M)\,
  dM_{n,N}
\end{equation}
holds for all functions $M \mapsto F(M)$ on $\mathrm{Herm}^+
(\mathbb{C}^n)$ with finite integral $\int F(\varphi^\ast \varphi)
\, d\varphi \, d\bar\varphi$.  In other words, $\psi$ pushes the
measure $d\varphi\, d\bar\varphi$ forward to
\begin{displaymath}
  \psi(d\varphi \, d\bar\varphi) = \mathrm{Det}^{N-n} (M)\, dM_{n,N}
    \;.
\end{displaymath}
While this claim can be viewed and proved as a statement in
invariant theory, the most elementary proof is to express the
integrals on both sides in terms of generalized polar coordinates,
as follows.

Given any complex rectangular matrix $\varphi \in \mathrm{Hom}
(\mathbb{C}^n , \mathbb{C}^N)$ for $N \ge n$, consider the
non-negative Hermitian matrices $M = \varphi^\ast \varphi$ and
$M^\prime = \varphi \, \varphi^\ast$, which are of size $n \times n$
and $N \times N$ respectively. The rank of $M^\prime$ cannot exceed
$n$, so there must be at least $N - n$ zero eigenvalues.  The other
$n$ eigenvalues are in general non-zero, and coincide with the
eigenvalues of $M = \varphi^\ast \varphi$. Denote these eigenvalues by
$\lambda_1, \ldots, \lambda_n$; their positive square roots
$\sqrt{\lambda_k}$ are sometimes called the \emph{singular values} of
$\varphi$.  There always exist two unitary matrices $U \in {\rm U}(n)$
and $V \in {\rm U}(N)$ such that
\begin{displaymath}
  \varphi^\ast = U \sqrt{\lambda}^{\rm T} V^{-1} \;, \quad \varphi = V
  \sqrt{\lambda} \, U^{-1} \;,
\end{displaymath}
where $\sqrt{\lambda}$ is the rectangular $N \times n$ matrix with
diagonal entries $\sqrt{\lambda_1}, \ldots, \sqrt{\lambda_n}$ and
zeroes everywhere else.

Let $J(\sqrt{\lambda})$ be the Jacobian of this singular value (or
polar) decomposition:
\begin{displaymath}
  J(\sqrt{\lambda}) = \prod_{1 \le i < i^\prime \le n} \Big(
  \sqrt{\lambda_i} - \sqrt{\lambda_{i^\prime}}\, \Big)^2 \Big(
  \sqrt{\lambda_i} + \sqrt{ \lambda_{i^\prime}}\, \Big)^2
  \prod_{k = 1}^n \sqrt{\lambda_k}^{1 + 2(N-n)} \;.
\end{displaymath}
Fix the values of $n$ and $N \ge n$. Then by a standard argument
there exists some (fixed) choice of Haar measure $dU$ for
$\mathrm{U}(n)$ such that
\begin{displaymath}
  \int\limits_{\mathrm{Hom}(\mathbb{C}^n, \mathbb{C}^N)} F(
  \varphi^\ast \varphi ) \, d\varphi \, d\bar\varphi =
  \int_{\mathbb{R}_+^n} \left( \int_{\mathrm{U}(n)} F(U \lambda
  U^{-1}) \, dU \right) J(\sqrt{\lambda}) \prod_{k = 1}^n
  d\sqrt{\lambda_k}
\end{displaymath}
holds for all integrable $F$.  Here $\lambda = \mathrm{diag}(
\lambda_1, \lambda_2, \ldots, \lambda_n)$.

On the other hand, fix some Lebesgue measure $dM$ for $\mathrm{
Herm} (\mathbb{C}^n)$.  By diagonalizing the Hermitian matrix $M$
and transforming the integral $\int f(M) dM$ to the eigenvalue
representation $M = U \lambda U^{-1}$ you get
\begin{displaymath}
  \int\limits_{\mathrm{Herm}^+(\mathbb{C}^n)} f(M) \, dM = b_{n,N}
  \int_{\mathbb{R}_+^n} \left( \int_{\mathrm{U}(n)} f(U \lambda
  U^{-1}) dU \right) \prod_{i < j} (\lambda_i - \lambda_j)^2 \prod_k
  d\lambda_k \,.
\end{displaymath}
The constant $b_{n,N}$ is determined by the (arbitrary) choice of $dM$
relative to $dU$.  Now put $f(M) = \mathrm{Det}^{N - n}(M)\, F(M)$.
Since $\mathrm{Det}^{N - n}(M)= \prod_k \lambda_k^{N - n}$, the
desired statement (\ref{bosonize}) follows (with $c_{n,N} = 2^{-n} /
b_{n,N}$) by comparing expressions and noting
\begin{displaymath}
  \prod\nolimits_{i < j} (\lambda_i - \lambda_j)^2 \prod\nolimits_k
  \lambda_k^{N - n} d\lambda_k = 2^n J(\sqrt{\lambda})
  \prod\nolimits_k d \sqrt{\lambda_k} \;.
\end{displaymath}

The relation (\ref{bosonize}) can also be viewed from another
perspective, which we shall now offer.  First note that the integral
on the left-hand side can be regarded as a distribution (or continuous
linear functional), say $\mu$, on $\mathrm{Herm}^+ (\mathbb{C}^n)$:
\begin{displaymath}
  \mu : F \mapsto \int F( \varphi^\ast \varphi ) \, d\varphi \,
    d\bar\varphi \;.
\end{displaymath}
Next observe that the non-compact Lie group $\mathrm{GL}(n, \mathbb
{C})$ acts transitively on the positive Hermitian $n \times n$
matrices $M$ by
\begin{displaymath}
  M \mapsto T M T^\ast \quad (T \in \mathrm{GL}(n,\mathbb{C})) \;.
\end{displaymath}
Via this action we can identify (a dense open subset of) $\mathrm{
Herm}^+ (\mathbb{C}^n)$ with the non-compact symmetric space $\mathrm
{GL}(n,\mathbb{C}) / \mathrm{U}(n)$.  The corresponding action on
functions, $F \mapsto {}^T F$, is given by
\begin{displaymath}
  {}^T F (M) = F \big( T^{-1} M {T^{-1}}^\ast \big) \;.
\end{displaymath}
Given that $T \in \mathrm{GL}(n,\mathbb{C})$ acts on the functions,
there is also an action $\mu \mapsto T(\mu)$ on the distributions, by
$T(\mu)[F] = \mu[ {}^{T^{-1}} F ]$.  Since the Jacobian of the
transformation $\varphi \mapsto \varphi \circ T^\ast$ and $\varphi
^\ast \mapsto T \circ \varphi^\ast$ is $\mathrm{Det}^N (T^\ast T)$,
the distribution $\mu$ at hand satisfies
\begin{displaymath}
  T(\mu) = \mathrm{Det}^N (T^\ast T) \,\, \mu \;.
\end{displaymath}

Now we make the same considerations on the right-hand side of
(\ref{bosonize}), i.e.\ for the distribution
\begin{displaymath}
    \widetilde{\mu} : F \mapsto \int F(M)\, \mathrm{Det}^{N-n}
    (M) \, dM \;.
\end{displaymath}
Under the transformation $M \mapsto T M T^\ast$ the Lebesgue measure
$dM$ transforms into $\mathrm{Det}^n (T^\ast T) \, dM$.  Hence
$\mathrm{Det}^{-n}(M) \, dM$ is invariant under such transformations,
and
\begin{displaymath}
  T(\widetilde{\mu}) = \mathrm{Det}^N (T^\ast T) \,\,
  \widetilde{\mu} \;,
\end{displaymath}
i.e., $\widetilde{\mu}$ transforms in exactly the same way as
$\mu$. It is an invariant-theoretic fact --- resulting from the
interpretation of $\mu$ as a linear functional on $\mathrm{U}
(N)$-invariant state vectors $F$ in the oscillator or Shale-Weil
representation of $\mathrm{Sp}(2nN)$ --- that the vector space of
distributions with this transformation property has dimension
one. Therefore, there exists some constant $c_{n,N}$ such that
\begin{displaymath}
  \mu = c_{n,N} \times \widetilde{\mu} \;.
\end{displaymath}

As a corollary, we note that $\mathrm{Det}^{-n}(M)\, dM$ is an
invariant measure for the symmetric space of positive Hermitian $n
\times n$ matrices, $\mathrm{Herm}^+ ( \mathbb{C}^n) \simeq
\mathrm{GL}(n, \mathbb{C}) / \mathrm{U}(n)$. The case encountered
in the main text of the paper is obtained by replacing $n \to 2n$.

\end{document}